\documentclass[a4paper,12pt,onecolumn]{article}

\usepackage[left=1in,top=1in,right=1in,bottom=1in]{geometry}

\usepackage{graphicx}
\usepackage{natbib}

\usepackage{amsmath,amssymb,amsfonts}
\usepackage{color}
\usepackage{varioref}
\usepackage{wrapfig}
\usepackage[FIGTOPCAP, hang, raggedright, nooneline]{subfigure}
 \usepackage{subfigmat}
\usepackage{dcolumn}
\usepackage{bm}
\usepackage{multirow}
\usepackage{rotating}
\usepackage{lscape}

\usepackage{tikz}

\usepackage[T1]{fontenc}

\newsavebox{\astrutbox}
\sbox{\astrutbox}{\rule[-5pt]{0pt}{20pt}}

\title{The role of surface tension gradient in determining microscopic dynamic contact angle}

 \author{
  Joseph J. Thalakkottor\thanks{Department of Mechanical and Aerospace Engineering, University of Florida, Gainesville, FL 32611.}
  \ and Kamran Mohseni$^*$\thanks{Department of Electrical and Computer Engineering, University of Florida, Gainesville, FL 32611.}
  }

\begin{document}

\maketitle

\begin{abstract}
Following Gibb's interpretation of an interface as a dividing surface, we derive a model for the microscopic dynamic contact angle by writing a force balance for a control volume encompassing the interfaces and the contact line. In doing so we  identify that, in addition to the surface tension of respective interfaces, the gradient of surface tension plays an important role in determining the dynamic contact angle. This is because not only does it contribute towards an additional force, but it also accounts for the deviation of local surface tension from its static equilibrium value. It is shown that this gradient in surface tension can be attributed to the convective acceleration in the vicinity of the contact line, which in turn is a direct result of varying degree of slip in that region. In addition, we provide evidence that this gradient in surface tension is one of the key factors responsible for the difference in contact angle at the leading and trailing edge, of a steadily moving contact line. These findings are validated using molecular dynamics simulations.
\end{abstract}

\section{Introduction}

The phenomena of wetting plays an important role in many industrial and natural processes, such as liquid coating, printing, oil recovery \citep{JoshiH:15a,GarciaEJ:16a}, lubrication \citep{KumarG:09a,OkadaK:09a}, electrowetting \citep{Mohseni:07a} and microfluidics \citep{AhnDJ:08a}. The extent to which a fluid wets a solid surface is commonly characterized by the contact angle (angle formed by the fluid-fluid interface with the solid). The contact angle is critical in determining the shape of the interface, which is in fact what quantifies the wettability of a material system. The challenge with using contact angle for characterizing forced wetting is its measurement under dynamic conditions \citep{SnoeijerJH:13b,ChauTT:09a}, which is exacerbated by the lack of understanding and of a general consensus about the forces determining the shape of the interface near the moving contact line.

  \cite{YoungT:05a} determined the contact angle formed by a {\it static} droplet by balancing the forces acting at the contact point. His model related static contact angle to the surface tensions of the three media. However, application of Young's model requires knowledge of the fluid-wall surface tensions, which are not conveniently and reliably measured. \cite{DupreA:69a} eliminated this issue by introducing the ``work of adhesion'' to obtain what is commonly referred to as the Young-Dupr\'{e} equation. Even though over the years the validity of Young's equation has been challenged \citep{MakkonenL:16a}, it has withstood the test of time \citep{BlakeTD:17a} and is now widely accepted.
  
   The same cannot be said about the contact angle models in the {\it dynamic} case. There are several models in literature that are based on a diverse set of physical mechanisms. These models are often grouped into three categories based on their length scales. Here, we briefly mention some of the works under these categories, for a more detailed list we refer the reader to the following review articles, \cite{SnoeijerJH:13b,ChauTT:09a}. In the macro-scale, \cite{LandauLD:84b} predicted an apparent contact angle by extrapolating the interface profile  and \cite{CoxRG:86a} by using asymptotic expansion. In the meso-scale, the models were either based on hydrodynamic mechanisms \citep{VoinovOV:76a} or interface formation theory \citep{ShikhmurzaevYD:97a}. In the micro-scale, \cite{BlakeTD:95a} estimated the dynamic contact angle by accounting for an off-balance force acting in the contact line region (molecular kinetic theory, MKT). Although extensive work in this area has improved our understanding of the physics governing the shape of the interface and the contact angle, there is still a lack of consensus and understanding.


In this manuscript we systematically identify and validate the forces that determine the microscopic dynamic contact angle. This is achieved by writing a force balance for a control volume which encompasses the interfaces as well as the contact line \citep{SlatteryJC:07a}. The key finding is that surface tension gradient plays an important role in determining the microscopic dynamic contact angle. The gradient in surface tension not only results in the deviation of surface tension from its static value locally, but also contributes towards the force balance around the contact point. This gradient in surface tension exists irrespective of any temperature or concentration gradients along the interface, and is attributed to convective acceleration. Which in turn is a result of increasing velocity slip in the vicinity of a moving contact line \citep{Mohseni:16b,Mohseni:13g}. Additionally, surface tension gradient provide an explanation for the difference in contact angles at the leading and trailing edge of a steadily moving contact line. These findings are validated using molecular dynamics (MD) simulations for the case of forced wetting.
 
 \section{Deriving the microscopic contact angle model}\label{sec:model}
 Before we delve into writing the force balance for a control volume encompassing the interfaces and the contact line, we first revisit the concept of an interface. In reality, the interface between two phases of matter has a finite thickness across which physical quantities vary continuously. However, the continuum assumption often interprets an interface  as a mathematical line where properties change discontinuously. \cite{GibbsJB:28a} on the other hand refers to the interface as a dividing surface. He defines it as the surface that separates a system into two homogeneous bulk phases with excess quantities in the interface region assigned to the dividing surface. The intersection of two or more such dividing surfaces result in a geometric and physical irregularity, and contact line is an example of it. In this manuscript we follow Gibb's interpretation of an interface.
 
  \begin{figure}
   \begin{minipage}{0.43\linewidth}\raggedright (a)\end{minipage}
    \begin{minipage}{0.59\linewidth}\raggedright (b)\end{minipage}\\
 \begin{minipage}{0.43\linewidth}
 \centerline{
  \includegraphics[width=0.85\textwidth]{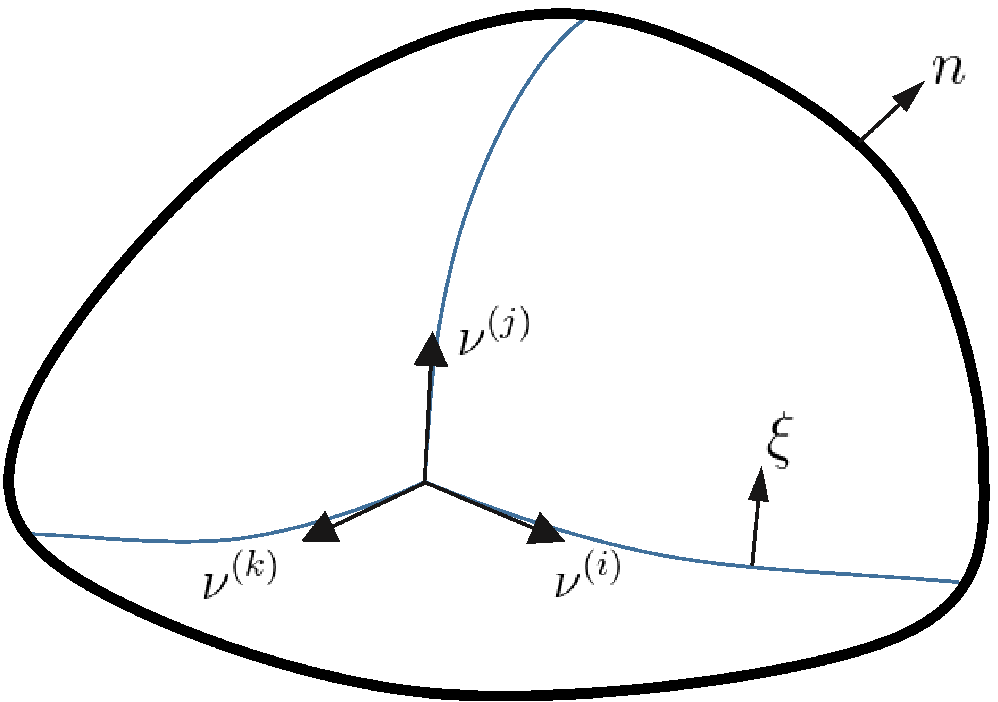}}
  \end{minipage}
 \begin{minipage}{0.59\linewidth}
 \centerline{
  \includegraphics[width=0.85\textwidth]{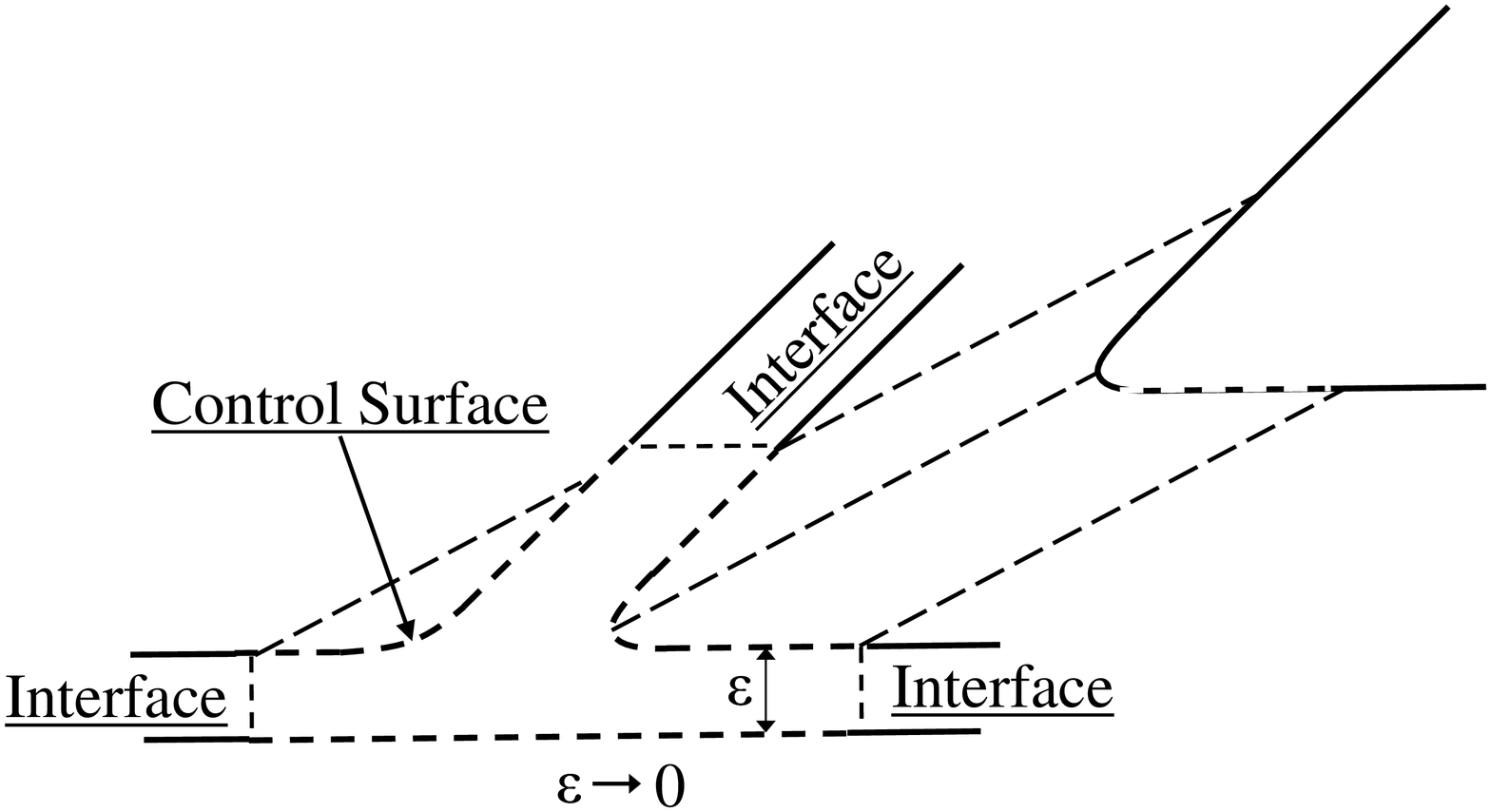}}
 \end{minipage}
   \caption{(a) Schematic of an arbitrary control volume encompassing a region $R$. $\bf{n}$ is the unit vector normal to the control surface, $\bf{\xi}$ is the unit vector normal to the dividing surface and $\bf{\nu}$ is the unit vector tangent to individual interfaces at the contact line. The inset shows that at molecular scales the interfaces and their intersection have a finite thicknesses. (b) Schematic of the control surface for the specific case of a contact line problem.}
 \label{fig:Schematic-CV}
 \end{figure}
  
We start by writing the conservation of momentum for a general multiphase system, as presented by \cite{SlatteryJC:07a}. Following which, we simplify it for the simple case of a contact line steadily moving on a planar wall. Taking the component of force balance along the planar wall, we obtain a model for the microscopic dynamic contact angle. The key difference in our approach from that commonly seen in literature, is in our choice of control volume over which the momentum is balanced. The conventional momentum balance for multi-phase systems typically writes separate balances for each individual phase and imposes matching conditions at the respective interfaces. However, by conserving the momentum in a control volume (figure \ref{fig:Schematic-CV}) encompassing the three phases of matter, the three interfaces and the contact line, the need for separate matching conditions at the interface is eliminated. In fact, such conditions become an output of the balance. This approach ensures a balance of momentum in the bulk media, the interface and the contact line. Hence, we start by writing the conservation of momentum for a general multiphase system, which is given as
 \begin{align}
 \label{eq:general_mom}
 &\overbrace{\int_{\text{R}} \left(\rho \frac{d_{(\text{m})}{\bf v}}{dt} - \nabla\cdot {\bf T} - \rho {\bf b} \right)dV}^{\text{momentum balance in the bulk media}}\\ \notag
 &+\overbrace{\int_{\Sigma} \left(\rho^{(\text{a})}\frac{d_{(\text{a})}  {\bf v}^{(\text{a})}}{dt} -  \nabla_{(\text{a})}\cdot {\bf T}^{(\text{a})} - \rho^{(\text{a})} {\bf b}^{(\text{a})}
 +\left[\!\!\left[\rho ({\bf v} - {\bf v}^{(a)})({\bf v} - {\bf v}^{(a)})\cdot{\bm \xi} - {\bf T}\cdot{\bm \xi} \right]\!\!\right]\right) dA}^{\text{momentum balance in the interface}}\\ \notag
 &+\overbrace{\int_{\text{c}}\left(\rho^{(\text{l})}\frac{d_{(\text{l})}  {\bf v}^{(\text{l})}}{dt} - \nabla_{(\text{l})}\cdot {\bf T}^{(\text{l})} - \rho^{(\text{l})} {\bf b}^{(\text{l})}
 + \left[\!\!\left[\rho^{(\text{a})}({\bf v}^{(\text{a})} - {\bf v}^{(\text{l})})({\bf v}^{(\text{a})} - {\bf v}^{(\text{l})})\cdot {\bm \nu} - {\bf T}^{(\text{a})}\cdot{\bm \nu} \right]\!\!\right]\right)ds}^{\text{momentum balance in the contact line}}\\ \notag
 &=0.
 \end{align}
 Here, $\rho$ is the density, ${\bf v}$ is the velocity, ${\bf T}$ is the stress tensor, and ${\bf b}$ is the body force. Terms without any subscript or superscript are associated with bulk media, $(\cdot)_{(\text{a})}$ and $(\cdot)^{(\text{a})}$ are terms associated with an interface, while $(\cdot)_{(\text{l})}$ and $(\cdot)^{(\text{l})}$ are associated with the contact line. $R$ is the region encompassed by the control volume, $\Sigma$ is the area formed by the interface and $c$ is the contact line (normal to the plane of the paper). The terms have been grouped into volume, area, and line integrals. The volume integral is the same as the commonly used momentum equation for a bulk medium. The surface and line integrals are similar to the bulk equation of momentum, except for the additional jump terms denoted by $\left[\!\!\left[\cdot\right]\!\!\right]$.
 The first of the jump terms accounts for the mass transfer ($\left[\!\!\left[\rho ({\bf v} - {\bf v}^{(\cdot)})({\bf v} - {\bf v}^{(\cdot)})\right]\!\!\right]$) across the dividing surfaces and their intersections, and the second accounts for the jump in stress normal ($\left[\!\!\left[{\bf T}\cdot{\bm \xi}\right]\!\!\right]$, $\left[\!\!\left[{\bf T}^{(\text{a})}\cdot{\bm \nu}\right]\!\!\right]$) to them. 
 It must be noted that this is a general momentum equation or force balance and as such is not limited to any specific problem. Using the above equation \cite{MohseniK:16x} studied the shedding of a vortex sheet at the trailing edge of an airfoil, where the vortex sheet is considered as a dividing surface.
 
The equation for the conservation of momentum of a general multiphase system is simplified by considering a simple contact line steadily moving on a planar wall. Consequently, the following assumptions are made:
\begin{enumerate}
 \item the system is in dynamic equilibrium, $\partial (.)/\partial t = 0$, 
 \item it is subject to no body force, $b^{(.)}=0$, 
 \item the interface are material boundaries, $\left[\!\!\left[\rho ({\bf v}^{(.)} - {\bf v}^{(.)})({\bf v}^{(.)} - {\bf v}^{(.)})\cdot{\bf \xi} \right]\!\!\right] = 0$, 
 \item there is no gradient in line stress along the contact line (normal to the plane of the paper) $\nabla_{(\text{l})}\cdot {\bf T}^{(\text{l})}=0$ and
 \item the momentum is conserved in the bulk media, $\nabla\cdot {\bf T}=0$.
\end{enumerate}
%
 Furthermore, for a general case, surface stress can be described by the Boussinesq surface fluid model \citep{Boussinesq:1913a}, ${\bf T^{(\text{a})}} = \left[\gamma + (\kappa - \epsilon){\bf \nabla_{(a)}\cdot v^{(a)}}\right]{\bf P} + 2\epsilon{\bf D}^{(a)}$. Here, $\kappa$ is the surface dilatational viscosity, $\epsilon$ is the surface shear viscosity, ${\bf P}$ is the surface projection tensor and ${\bf D}^{(a)}$ surface rate of deformation tensor. If we assume: (i) the domain to be periodic in the $z$ direction, ${\bf D}^{(a)}=0$, and (ii) the surface flow to be incompressible, ${\bf \nabla_{(a)}\cdot v^{(a)}}=0$,
then the surface stress can be written as ${\bf T^{(\text{a})}} = \gamma$.
 Hence, for the specific case of a contact line steadily moving over a non-porous planar wall, with no temperature or concentration gradients, the momentum balance reduces to
 \begin{equation}
 \int_{\Sigma} \left( \rho_a {\bf v^{(\text{a})}\nabla_{(\text{a})}\cdot v^{(\text{a})}} +  \nabla_{(\text{a})}\gamma  +\left[\!\!\left[  {\bf T}\cdot{\bm \xi} \right]\!\!\right]\right) dA
 +\int_{\text{c}}\left[\!\!\left[  \gamma{\bm \nu} \right]\!\!\right]ds
 =0.
 \label{eq:contact_model}
 \end{equation}
The forces exerted on the interface (dividing surface) are due to interface inertia, surface gradient of surface tension and the jump in bulk stress tensor normal to the interface. 
The force exerted at the contact line is due to the jump in surface tension  $\left( \int_{\text{c}} \left[\!\!\left[  \gamma{\bm \nu} \right]\!\!\right]ds\right)$. This jump represents the balance of surface tension forces at the contact line, which is the well-known Young's equation for a static contact angle. 

 Now, by taking the component of force along the wall and re-writing it for a 2D problem, we obtain
  \begin{align}
 \label{eq:general_mom_2D}
 \cos \theta \left[\int \nabla_{(\text{a})} { \gamma_{AB}} dl
 +
 {\gamma_{AB}}\right]
 =\int \left(\nabla_{(\text{a})} { \gamma_{AW}}+\nabla_{(\text{a})}{ \gamma_{BW}}
\right) dl
 + \left({\gamma_{AW}} +{\gamma_{BW}} \right).
 \end{align}
Here the subscripts $AB$, $AW$ and $BW$ stand for fluid A-fluid B, fluid A-wall and fluid B-wall interfaces, respectively.
Above, we have assumed the contributions from forces due to inertia and the jump in stress across the fluid-fluid interfaces to be negligible in comparison to surface tension forces. This is confirmed on looking at the respective force contributions in Appendix A. As for the fluid-wall interface, there is no jump in stress across it.
Hence, if we want to evaluate the microscopic contact angle at an infinitesimal distance away from the contact point, the model (\ref{eq:general_mom_2D}) shows that for the dynamic case, in addition to the surface tension of respective interfaces, we also need to account for the force due to the gradient of surface tension. The surface tension gradient plays a key role as not only does it contribute towards the force balance but it also causes deviation of the local value of the surface tension from its equilibrium static value. In the static limit the contribution from the force due to surface tension gradient reduces to zero. The model is then reduced to the Young's equation. The presence of surface tension gradient, near the moving contact line, and the cause of it is discussed in detail in section \ref{sec:results}. Now, if we want to evaluate the microscopic contact angle at the limit of the moving contact point, the model reduces to a force balance of just the local surface tension forces same as that of Young's equation,
   \begin{equation}
 \label{eq:general_mom_2D_at_CP}
 \cos \theta {\gamma_{AB}}
 = {\gamma_{AW}} +{\gamma_{BW}} .
 \end{equation}
 Although, it must be noted that as a result of surface tension gradient in the vicinity of the moving contact point, the local surface tension value deviates from the static equilibrium value. This is consistent with \cite{BlakeTD:17b}, where they also show that in the limit of the moving contact point, the force balance is given by Young's equation.
 
To summarize this section, by systematically deriving the model for microscopic contact angle we are able to decompose and identify the forces determining the dynamic contact angle. 
It must be noted that by using the appropriate slip boundary condition, one can evaluate the velocity field in the domain and hence evaluate strain rates and fluid stresses. The fluid stresses in turn can be used to evaluate the local surface tension, as per the mechanical definition.
Hence, the dynamic contact angle becomes part of the solution and is not required to be prescribed a priori. In the following sections we validate this model using molecular dynamic simulations.
 
 
 \section{Numerical setup and force evaluation}\label{sec:types_paper}
 Here, we describe the problem geometry, provide details of the numerical simulation and describe the method used to evaluate physical quantities from MD simulations.
 \subsection{Description of problem geometry}
\begin{figure}
   \begin{minipage}{0.89\linewidth}\raggedright (a)\end{minipage}\\
 \begin{minipage}{0.89\linewidth}
 \centerline{
  \includegraphics[width=0.85\textwidth]{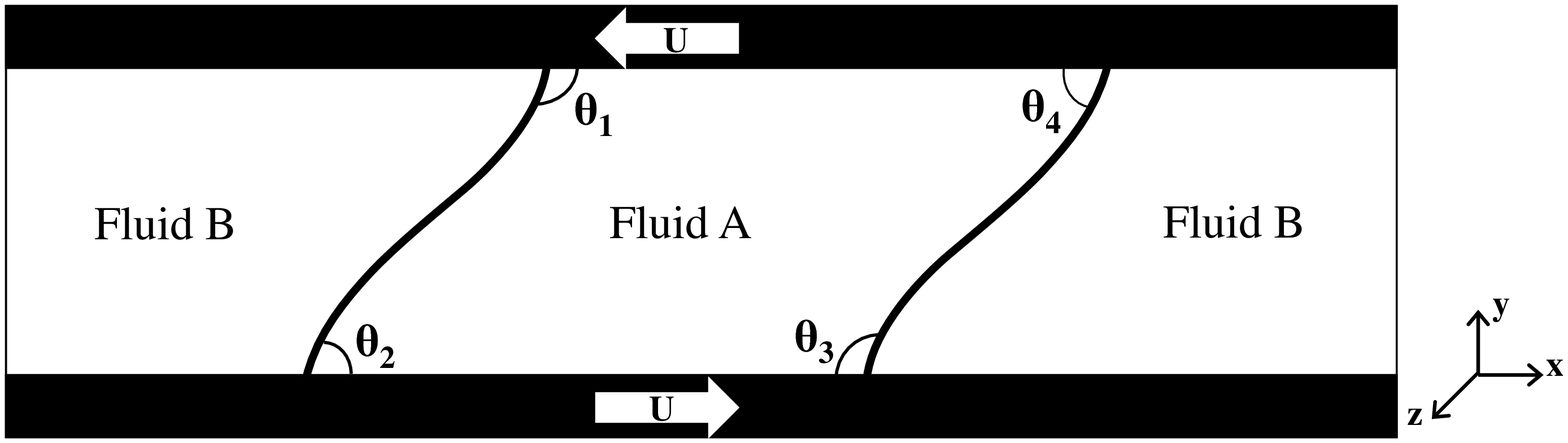}}
  \end{minipage}
 \begin{minipage}{0.89\linewidth}\raggedright (b)\end{minipage}\\
 \begin{minipage}{0.89\linewidth}
 \centerline{
  \includegraphics[width=0.85\textwidth]{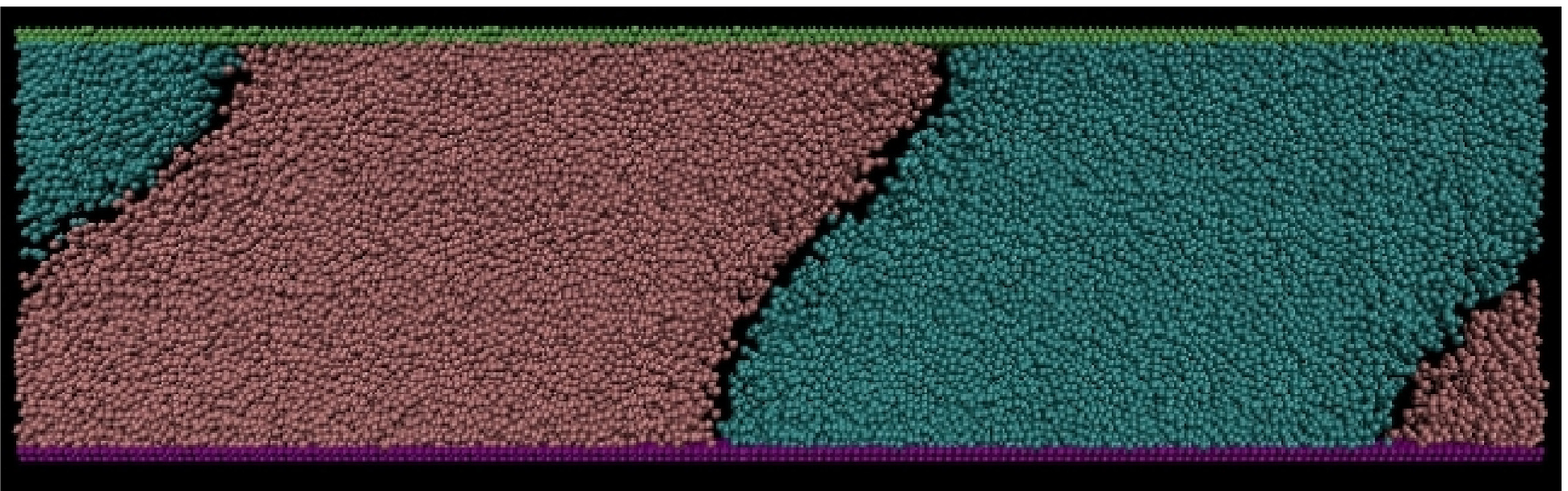}}
 \end{minipage}
   \caption{(a) Schematic of the problem geometry and (b) snapshot of the simulation.}
 \label{fig:ProblemGeometry}
 \end{figure}

 The moving contact line is simulated by modeling a two-phase Couette flow, where the walls move in opposite directions with a constant speed $U$, see figure \ref{fig:ProblemGeometry}.  Here, periodic boundary conditions are imposed along the $x$ and $z$ directions. The contact angle ($\theta$) is defined as the angle formed by the wall and fluid-fluid interface, as measured in fluid A. Due to the symmetric nature of the problem $\theta_1=\theta_3$ and $\theta_2=\theta_4$.
 
 \subsection{Details of MD simulation}
  As previously eluded to, one of the major difficulties faced in studying the dynamic contact angle is that direct measurement is limited by the ability of an experimental technique to resolve the interface. With the aide of molecular dynamics simulations \citep{PlimptonS:95a} we are able to overcome this hurdle. The pairwise interaction of molecules, separated by a distance $r$, is modeled by the Lennard-Jones (LJ) potential
 \begin{equation}
 V^{LJ}=4\epsilon\left[\left(\frac{\sigma}{r}\right)^{12}-\left(\frac{\sigma}{r}\right)^{6}\right].
 \end{equation}
 Here, $\epsilon$ and $\sigma$  are the characteristic energy and length scales, respectively. The potential is zero for $r>r_c$, where $r_c$ is the cutoff radius, which we set to $r_c=2.5\sigma$, unless otherwise specified. 
 
 Each wall is comprised of at least two layers of molecules oriented along the $(111)$ plane of a face centered cubic (fcc) lattice, with the molecules fixed to their respective lattice sites. The fluid molecules are initialized on a fcc lattice whose spacing is chosen to obtain the desired density, with initial velocities randomly assigned so as to obtain the required temperature. The fluid in its equilibrium state has a temperature $T\approx1.1k_B/\epsilon$ and number density $\rho\approx0.81\sigma^{-3}$. The temperature is maintained  using a Langevin thermostat with a damping coefficient of $\Gamma=0.1\tau^{-1}$, where $\tau=\sqrt{m\sigma^{2}/\epsilon}$ is the characteristic time and $m$ is the mass of the fluid molecule. The damping term is only applied to the $z$ direction to avoid biasing the flow.  The equation of motion of a fluid atom of mass $m$ along the $z$ component is therefore given as follows
 \begin{equation}
  m\ddot{z_i}=\sum_{j\neq i}\frac{\partial V{ij}}{\partial z{i}}-m\Gamma \dot{z_i} + \eta_i.
 \end{equation}
 Here $\sum_{j\neq i}$ denotes the sum over all interactions and $\eta_i$ is a Gaussian distributed random force. The value of dynamic viscosity of the fluid is $\mu\approx2.0 \epsilon\tau\sigma^{-3}$ and the Reynolds number is $\text{Re}\approx1.0$.

 The LJ coefficients and relative density of the various cases simulated are listed in table \ref{tab:test_case}. The immiscibility of the two fluids is modeled by choosing appropriate LJ interaction parameters, such that the interatomic forces between them is predominantly repulsive. Since $\sigma^{ff}=3.0\sigma$, a cut-off radius of $r_c=5.0\sigma$ is used for the fluid A-fluid B interactions. For simplicity the two fluids are assigned identical fluid properties and only the properties of the fluid-wall interactions are changed. The fluid channel measures $153.0\sigma\times27.4\sigma\times144.0\sigma$.  
 \begin{table}
   \begin{center}
 \def~{\hphantom{0}}
   \begin{tabular}{cccccccccc}
        & \multicolumn{3}{c}{\underline{Wall-fluid A}} & \multicolumn{3}{c}{\underline{Wall-fluid B}} & \multicolumn{3}{c}{\underline{Fluid A-fluid B}}\\[3pt]
     Case & ~$\epsilon^{wf}/\epsilon$~ & ~$\sigma^{wf}/\sigma$~ & ~$\rho^*$~ & ~$\epsilon^{wf}/\epsilon$~ & ~$\sigma^{wf}/\sigma$~ & ~$\rho^*$~ & ~$\epsilon^{ff}/\epsilon$~ & ~$\sigma^{ff}/\sigma$~ & ~$\rho^*$~\\
       1~ &~$0.4$~ & ~$1.0$~ & ~$1.0$~  &~$0.6$~ & ~$1.0$~ & ~$1.0$~  &~$0.2$~ & ~$3.0$~ & ~$1.0$~\\ 
       2~ &~$0.2$~ & ~$1.0$~ & ~$1.0$~  &~$0.6$~ & ~$1.0$~ & ~$1.0$~   &~$0.2$~ & ~$3.0$~ & ~$1.0$~\\ 
       3~ &~$0.1$~ & ~$1.0$~ & ~$1.0$~  &~$1.0$~ & ~$1.0$~ & ~$1.0$~    &~$0.2$~ & ~$3.0$~ & ~$1.0$~\\
   \end{tabular}
   \caption{List of different test cases. Here, $\epsilon$ and $\sigma$  are the characteristic energy and length scales, respectively. $\rho^*$ is the relative density.}
   \label{tab:test_case}
   \end{center}
 \end{table}

 The equations of motion were integrated using the Verlet algorithm \citep{VerletL:67a,TildesleyDJ:87a} with a time step $\Delta t=0.002\tau$. The equilibration time is determined by the time it takes for the fluid-fluid surface tension to reach a steady value, which is $60000\tau$. The simulation is initially run until the flow equilibrates, after which spatial averaging is performed by dividing the fluid domain into rectangular bins of size $\sim0.52\sigma\times0.27 \sigma$ along the $x$--$y$ plane, and extending through the entire width of the channel. In addition to spatial averaging, time averaging is done for a duration of $20000\tau$ for the moving contact line problem.

 \subsection{Calculation of physical quantities from MD}
 In order to validate the contact angle model, we need to evaluate the respective forces and determine the microscopic contact angle from MD simulations. First, we start by evaluating the microscopic contact angle for which the shape of the interface needs to be defined.

\begin{figure}
   \begin{minipage}{0.49\linewidth}\raggedright (a)\end{minipage}
   \begin{minipage}{0.49\linewidth}\raggedright (b)\end{minipage}\\
 \begin{minipage}{0.49\linewidth}
 \centerline{
  \includegraphics[width=0.85\textwidth]{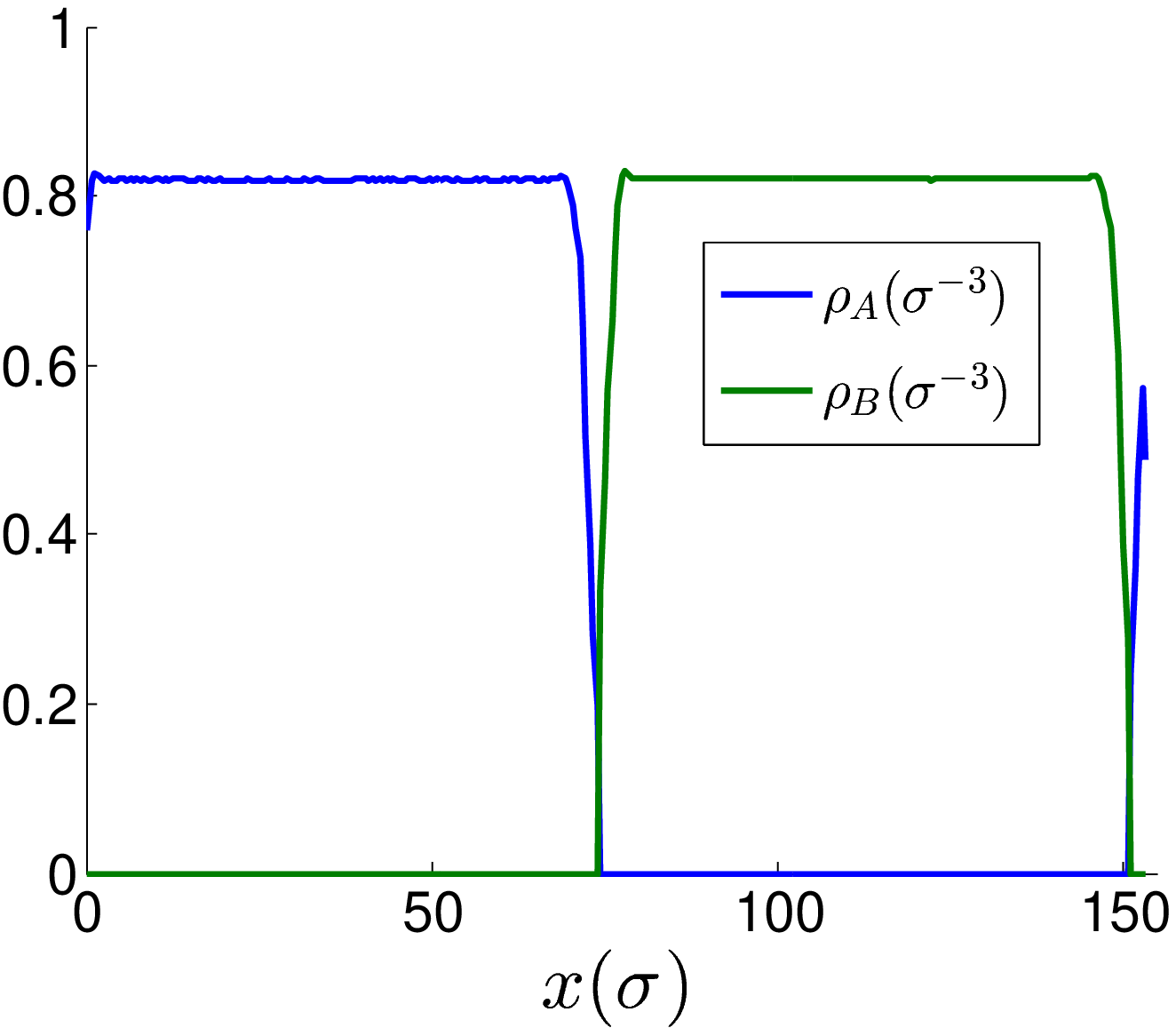}}
  \end{minipage}
 \begin{minipage}{0.49\linewidth}
 \centerline{
  \includegraphics[width=0.85\textwidth]{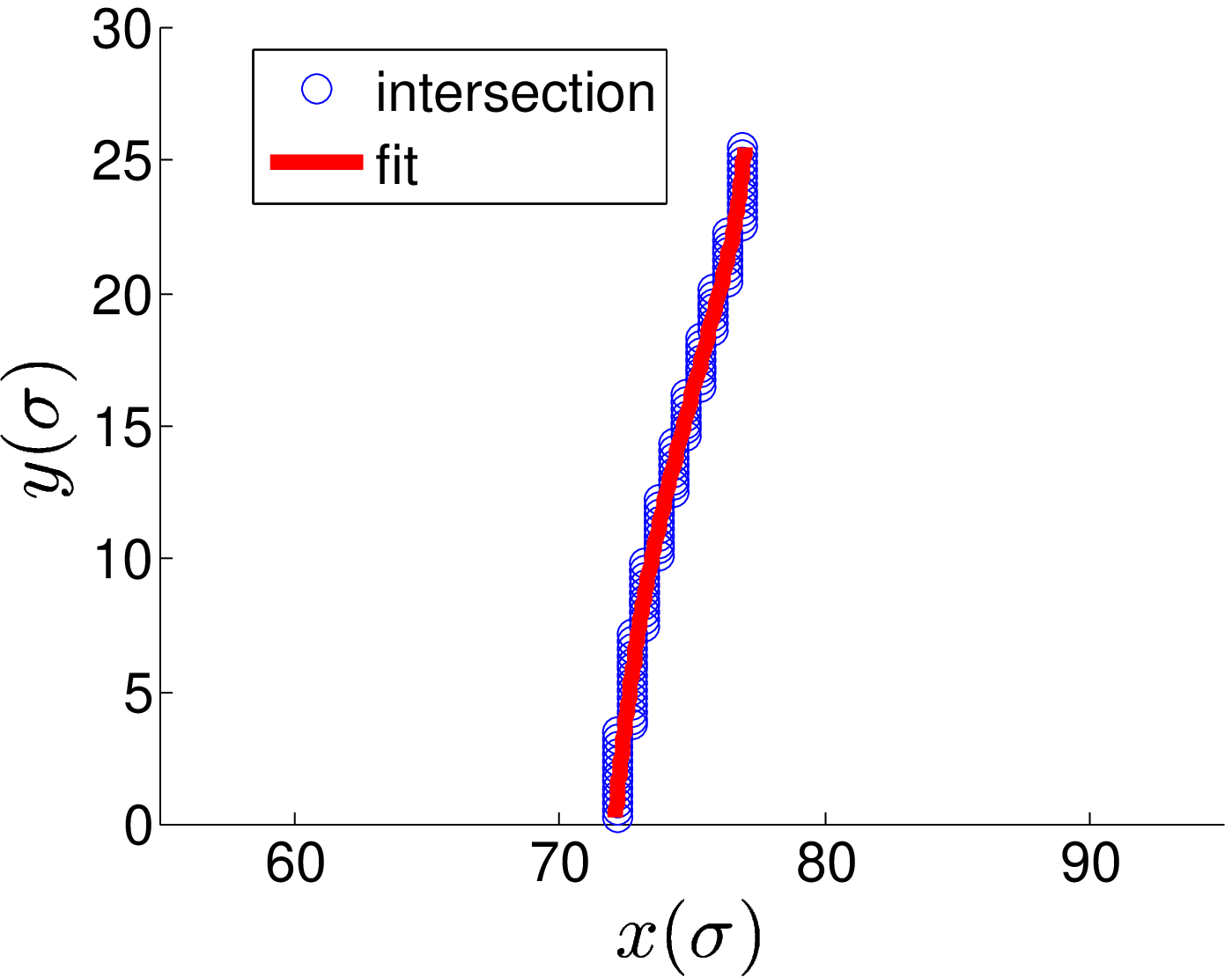}}
 \end{minipage}
   \caption{(a) Density profile of fluids A and B across the middle of the channel. (b) The polynomial fit to the location of all intersection points formed by the density profiles of fluids A and B.}
 \label{fig:Fig_IntfShapeFit}
 \end{figure}

 {\it Determining the microscopic contact angle:}
 From the density distribution along the length of the channel ($x$ direction), in figure \ref{fig:Fig_IntfShapeFit}(a), it is observed that the density of both fluids drop sharply within the interface region. We consider the location where the density of the two fluids intersect to lie on the fluid-fluid interface line. Hence, the fluid-fluid interface line is defined as the loci of the location of these intersections along the height of the channel, see figure \ref{fig:Fig_IntfShapeFit}(b). By fitting a cubic polynomial function through these points we obtain the approximate shape of the averaged fluid-fluid interface line. A cubic polynomial function is chosen to fit the data as the interface has two inflection points in this particular problem. Now that we have the function that describes the shape of the interface, we are able to calculate the contact angle by taking the tangent of this function, at any given distance away from the wall.
 Next, we define the control volume in which the respective forces in equation \ref{eq:general_mom_2D} are evaluated. In order to determine the control volume, we first need to define the interface and contact point region.

{\it Defining the interface region, the contact point region and the control volume:} 
 The width of the interface is determined by looking at the local density far away from the contact point. It is defined as the location where the local density deviates from the bulk value of fluid A and B, respectively. The threshold of this deviation is chosen to be $1\%$ of the bulk density. In the case of the wall-fluid interface the location of this deviation approximately corresponds to the height at which density layering in the fluid is observed.
 \begin{figure}
   \centerline{\includegraphics[width=0.85\textwidth]{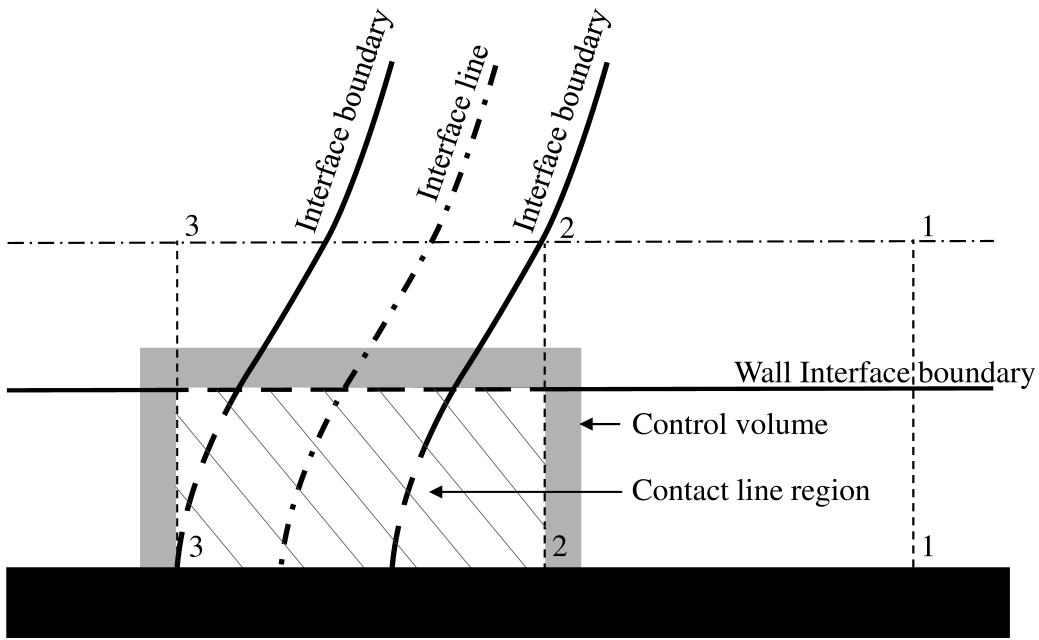}}
   \caption{Schematic showing how an approximate contact region is determined where $1-1$ is a location away from the interface. $2-2$ and $3-3$ are the locations of the right and left edges, respectively, of the contact region. The shaded region is the control volume with the hatched region which is the approximate contact region being eliminated from the control volume.}
 \label{fig:Schematic_intf_contact_region}
 \end{figure}
 The contact point region is defined as the intersection of the respective interfaces. Hence, the width and height of the contact point region is determined by the width of the fluid-fluid interface and the height of the fluid-wall interface. But, the extent to which we can accurately resolve this region is decided by our ability to evaluate the surface tension of a given interface without being affected by the neighboring interfaces. This is explained by referring to figure \ref{fig:Schematic_intf_contact_region}, where the difference in components of linear stress is integrated along the vertical line $1-1$, in order to evaluate the surface tension of the wall-fluid interface. The left and right boundaries of the contact line region are determined by where this line intersects with the boundaries of the fluid-fluid interface region (line $3-3$ for left boundary and line $2-2$ for right boundary). For the cases considered here, the contact point region was around $3\sigma\times4\sigma$, shown by the hatched area in figure \ref{fig:Schematic_intf_contact_region}.
 
Now, that we have defined the contact point region an infinitesimal control volume is defined around the contact point region. The smallest control volume that we can consider is determined by the minimum resolvable length scale. In our case the minimum that can be measured in a given direction is the bin width (bin height-$0.25\sigma$, bin width-$0.5\sigma$). Hence, the thickness of the control volume is $0.52\sigma$, depicted by the shaded region in the figure \ref{fig:Schematic_intf_contact_region}.
 

 {\it Evaluating the individual forcing terms:}
 Before we describe how the respective individual force terms are evaluated we refer back to Gibb's definition of dividing surface in order to evaluate various properties associated with the interface. As per Gibb's definition of a dividing surface, the excess quantities in the interface region are assigned to the dividing surface. Similarly, the excess quantities in the contact line region are assigned to the line formed by the intersection of dividing surfaces. 
 
\begin{figure}
   \begin{minipage}{0.49\linewidth}\raggedright (a)\end{minipage}
   \begin{minipage}{0.49\linewidth}\raggedright (b)\end{minipage}\\
 \begin{minipage}{0.49\linewidth}
 \centerline{
  \includegraphics[width=0.85\textwidth]{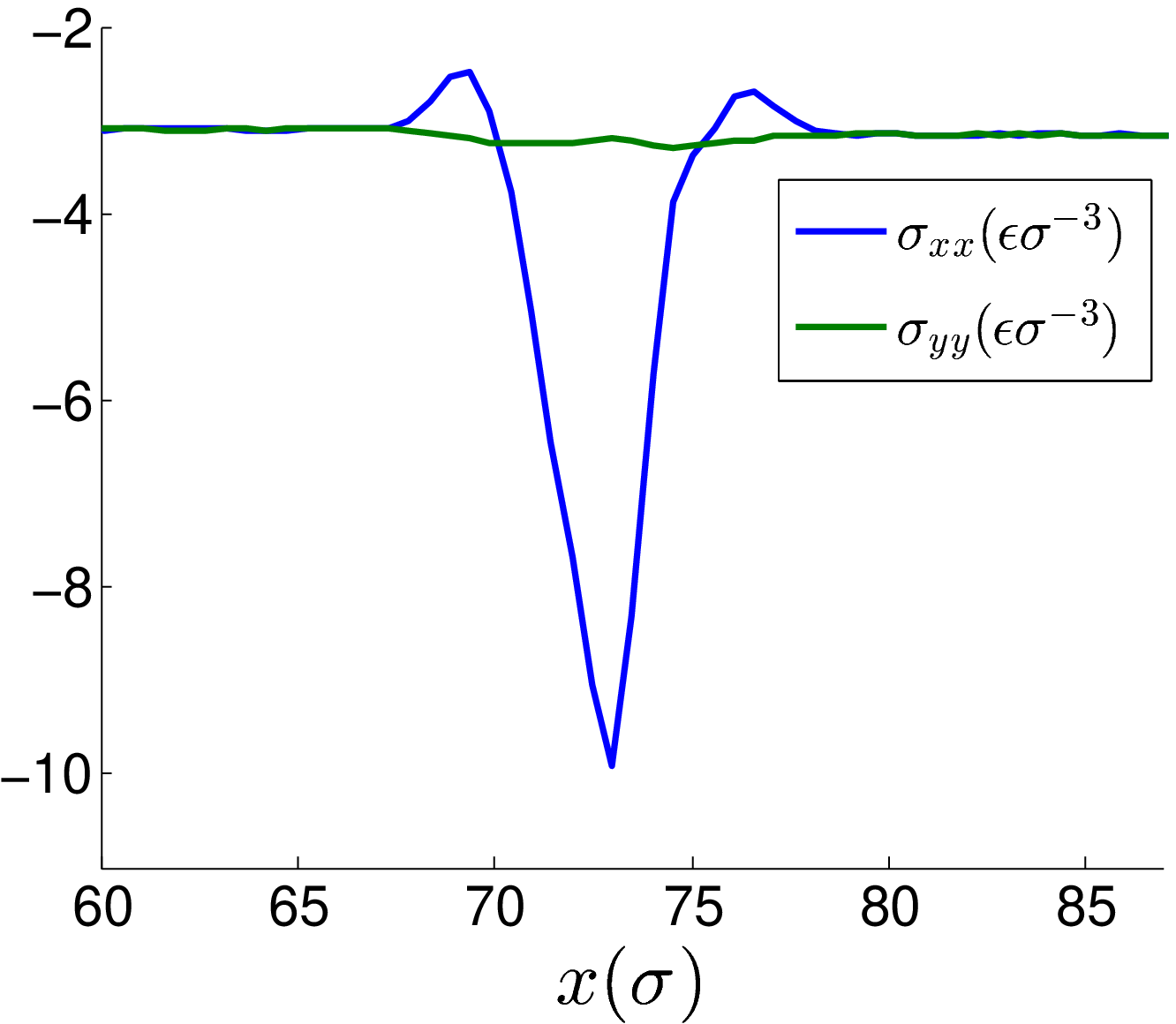}}
  \end{minipage}
 \begin{minipage}{0.49\linewidth}
 \centerline{
  \includegraphics[width=0.85\textwidth]{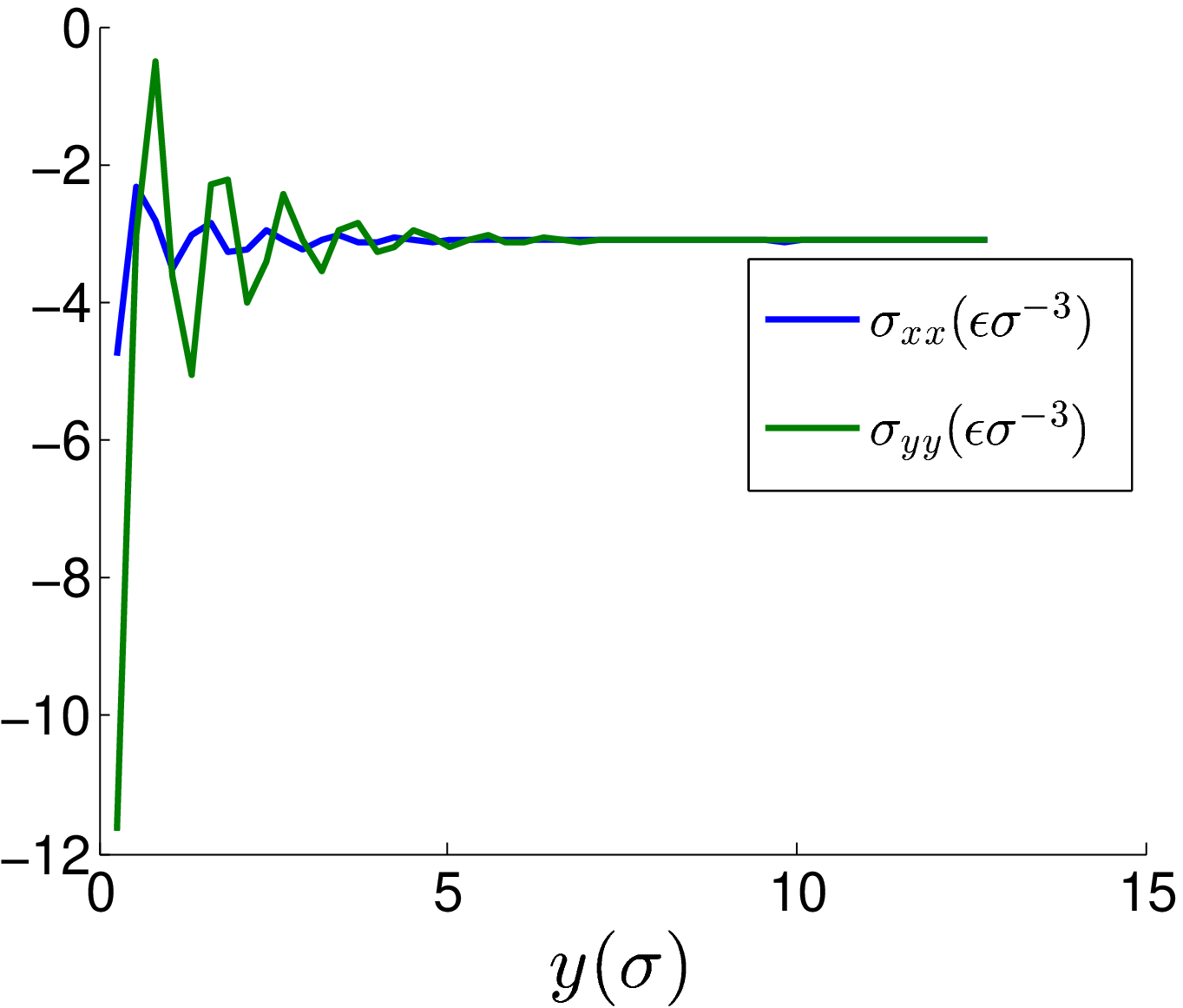}}
 \end{minipage}
   \caption{The components of linear stress in the $x$ and $y$ directions are plotted for (a) the fluid-fluid interface at the middle of the channel ($y=H/2$) and (b) the wall-fluid interface at the middle of fluid A ($x=78.394\sigma$). The results are for the static case corresponding to Case 1.}
 \label{fig:StressDistrib}
 \end{figure}

 The virial stress tensor is computed by accounting for the forces acting on it \citep{HeyesDM:94a,PlimptonSJ:09a,SirkTW:13a}. By spatially averaging the virial stress, we obtain the average stress tensor in each bin. In figure \ref{fig:StressDistrib}, we plot the stress components ($P_{xx}$ and $P_{yy}$) across the fluid-fluid and at the fluid-wall interfaces, where the anisotropy of stress around the interface is observed. Knowing the components of the stress tensor we now calculate the surface tension as given by its mechanical definition, $\gamma = \int_{L_1}^{L_2} [P_{N}-P_{T}]dl$ \citep{Kirkwood:48a}. Here, $P_N$ and $P_T$ are the normal and tangential components of the local stress tensor with respect to the interface and $dl$ is in a direction normal to the interface. The limits of integration are chosen such that at $l=L_1$ and $l=L_2$ the stress is isotropic ($P_N=P_T$).  In order to evaluate the surface tension for the fluid-fluid interface the limits $L_1=-5\sigma$ and $L_2=5\sigma$ are used. While evaluating close to the wall the limit only extends up to the wall. In the case of the wall-fluid interface, surface tension is computed using the same mechanical definition as given above since here we use inert walls \citep{NijmeijerMJP:90a}. The limits of integration for the wall-fluid interface extends from the wall to the middle of the channel. The gradients for the surface tension along the respective interface are evaluated using a central difference scheme and at the end points a one sided second order scheme. The force contribution due to surface tension is directly obtained by looking at the respective surface tensions in the control volume. In order to calculate the contribution due to the surface tension gradient, the value of this gradient is multiplied by the length of the control volume projected onto the interface. The force contribution corresponding to the normal stress across the interface is calculated in similar manner, but with the normal component of stress. 

%
%
  
 \section{Results and discussion}\label{sec:results}
In this section, we present the results from our MD simulations and validate our contact angle model. In addition, we discuss the importance of surface tension gradient in determining the dynamic contact angle and the cause of it.
 
{\it Validating contact angle model using MD results.} Now that we have described the problem geometry and provided the details of the numerical simulation, the microscopic contact angle model is validated by comparing the contact angles from several different cases of wall-fluid properties and for the varying wall velocities in table \ref{tab:contact_angle}. The prediction made by the microscopic dynamic contact angle model agrees well with the results of the MD simulations. For most cases (excluding case 3, $U=0.050$ and $U=0.075$) the error is under $5\%$. For the two excluded cases the increased error, of over $5\%$, corresponds to the extreme acute angles formed by the interface, where the current grid resolution limits our ability to accurately calculate the local surface tension near the contact point. For the limiting case of a stationary wall ($U=0$), the angle $\theta_1\approx\theta_2$ as dictated by Young's equation. This shows the consistency of both the MD simulation setup and the methodology used to evaluate individual force contributions. Recently, \cite{BlakeTD:17a} performed an extensive study using molecular dynamics simulations that validated the Young's equation for a static case. Here, even though the results are only presented for fluid A the model is also validated for fluid B, as they share the same interface. The contact angle for fluid B is $180-\theta$, where $\theta$ is the contact angle for fluid A. { The individual force contributions in eq. \ref{eq:contact_model} are tabulated for various test cases in Appendix A, table \ref{tab:force}}. It is shown that the force contribution due to the surface tension gradient along the solid-fluid interface is comparable to that of the surface tension, and hence cannot be neglected. The surface tension gradient not only results in an additional force, but it also alters the local surface tension value from its static equilibrium value.   As for the jump in stress normal to the interface ($|F_{\Delta P}|$) and the inertia term ($\rho_{AB}^{(\text{a})}{ v_{AB}^{(\text{a})} {\bf \nabla \cdot  v_{AB}^{(\text{a})}}}$), the force contribution is more than an order of magnitude smaller than that of the surface tension forces and validates are assumption of neglecting these terms. The effects and the generation of the surface tension gradient are discussed in detail below. 
 \begin{table}
   \begin{center}
 \def~{\hphantom{0}}
   \begin{tabular}{cccccccc}
       \multicolumn{2}{c}{} & \multicolumn{3}{c}{\underline{$\theta_1$}} & \multicolumn{3}{c}{\underline{$\theta_2$}}\\[3pt]
    Case &$U$ $\left(\sigma\tau^{-1}\right)$ &Sim (deg) &Mod (deg) &Err$(\%)$ &Sim (deg) &Mod (deg) &Err$(\%)$\\[3pt]
       1 & $0.050$~ &$104.34$~ & ~$107.99$~ & ~$3.50$~ & ~$81.34$~ & ~$84.09$~ & ~$3.38$~\\
       2 & $0.000$~ &$94.75$~ & ~$94.87$~ & ~$0.13$~ & ~$95.50$~ & ~$94.95$~ & ~$0.58$~\\
 	& $0.050$~ &$105.60$~ & ~$106.02$~ & ~$0.40$~ & ~$87.35$~ & ~$88.53$~ & ~$1.35$~\\
       3 & $0.000$~ &$104.33$~ & ~$107.80$~ & ~$3.33$~  & ~$103.55$~ & ~$108.30$~ & ~$4.59$~\\
 	& $0.025$~ &$110.79$~ & ~$109.19$~ & ~$1.44$~  & ~$96.15$~ & ~$103.02$~ & ~$7.14$~\\
 	& $0.050$~ &$118.91$~ & ~$110.95$~ & ~$6.69$~  & ~$87.35$~ & ~$98.31$~ & ~$12.55$~\\
 	& $0.075$~ &$131.00$~ & ~$112.01$~ & ~$14.50$~  & ~$81.04$~ & ~$95.35$~ & ~$17.66$~\\
   \end{tabular}
   \caption{The contact angle obtained from the MD simulation (Sim) is compared with that predicted by the microscopic contact angle model (Mod) for different test cases. Here the contact angles are measured with respect to fluid A and the absolute relative error is presented. Refer to schematic \ref{fig:ProblemGeometry} for the definitions of $\theta_1$ and $\theta_2$.}
   \label{tab:contact_angle}
   \end{center}
 \end{table}
 %

\begin{figure}
 \centering
 \begin{minipage}{0.49\linewidth}\raggedright (a)\end{minipage}
 \begin{minipage}{0.49\linewidth}\raggedright (b)\end{minipage}\\
 \begin{minipage}{0.49\linewidth}\begin{center}
  \includegraphics[width=1.0\textwidth]{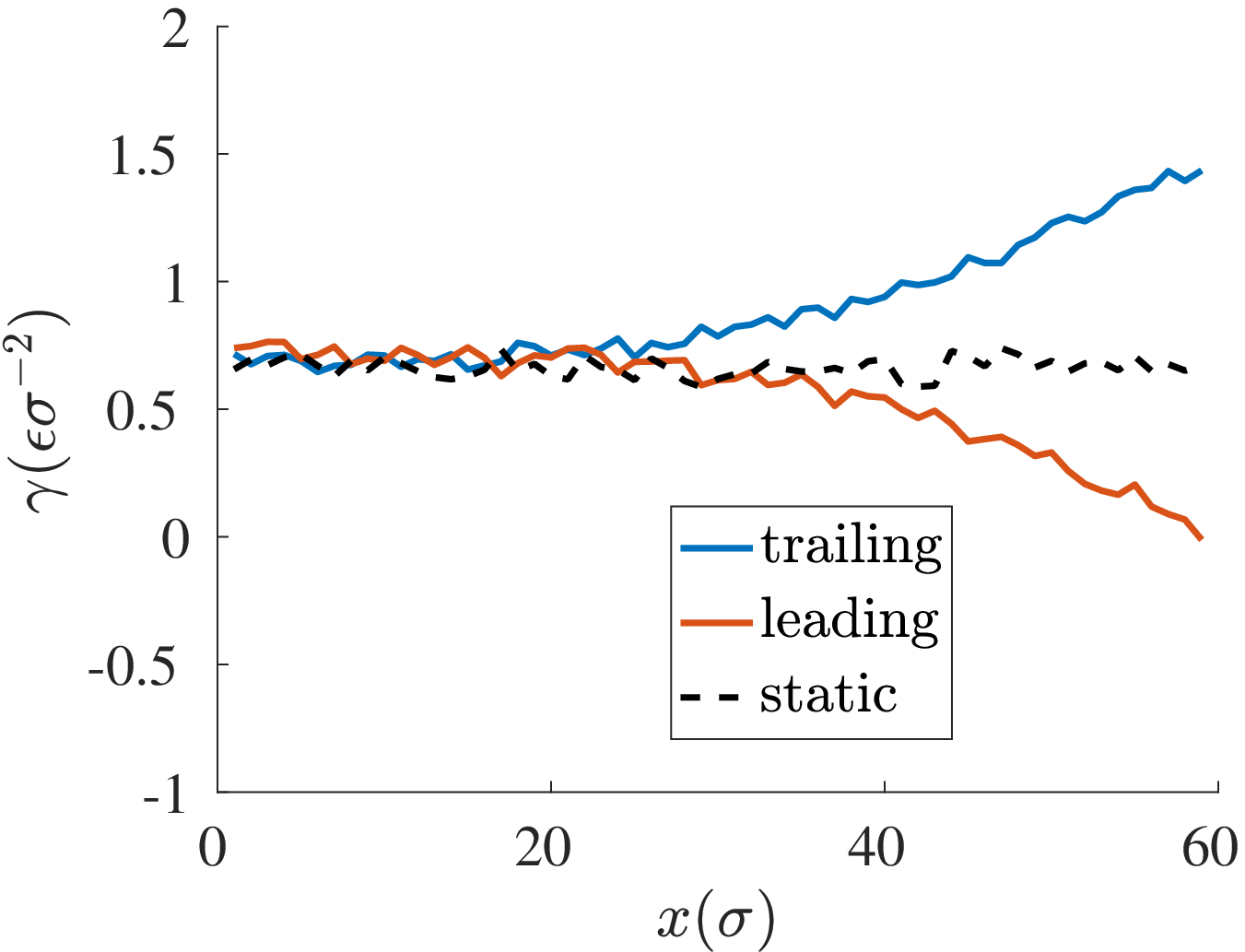}
 \end{center}
 \end{minipage}
 \begin{minipage}{0.49\linewidth}\begin{center}
  \includegraphics[width=1.0\textwidth]{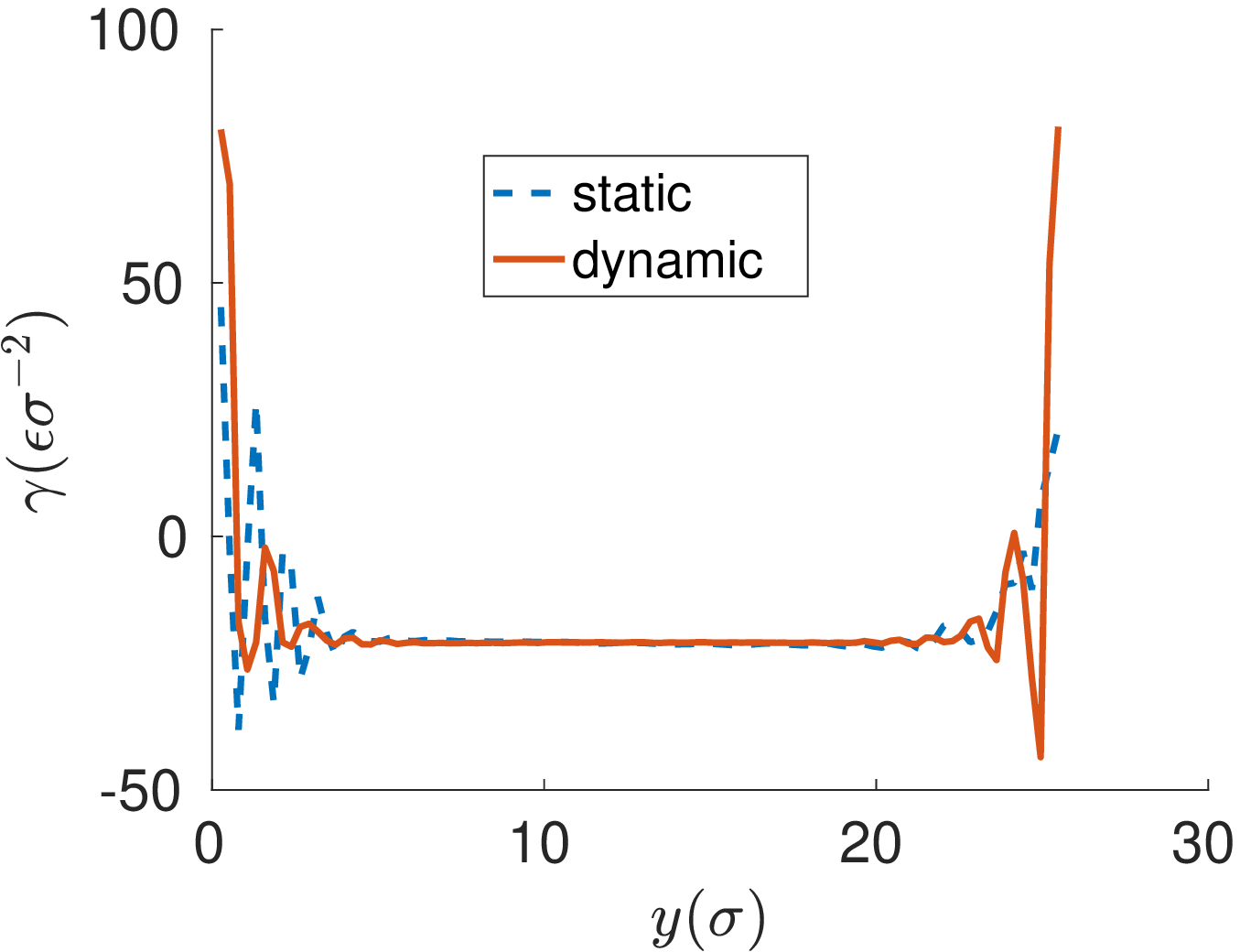}
 \end{center}
 \end{minipage}\\
 \caption{{Comparison of surface tension gradient for a static and dynamic case along the (a) wall and (b) fluid-fluid interface.} In figure (a) the results are shown for fluid B, whose interface happens to coincide with the start of the channel for this case. The oscillations in surface tension gradient along the $x$-axis for the static case are due to fluid structure induced by the wall in the adjacent fluid layers. This is alleviated by enhanced mixing in the dynamic case, in comparison to the static case. In figure (b) the oscillations near the wall are a result of the layering phenomenon \cite{ThompsonP:90a}.}
 \label{fig:SurfaceTension}
 \end{figure}

\begin{figure}
 \begin{minipage}{0.49\linewidth}\raggedright (a)\end{minipage}
 \begin{minipage}{0.49\linewidth}\raggedright (b)\end{minipage}\\
 \begin{minipage}{0.49\linewidth}
 \centerline{
  \includegraphics[width=0.85\textwidth]{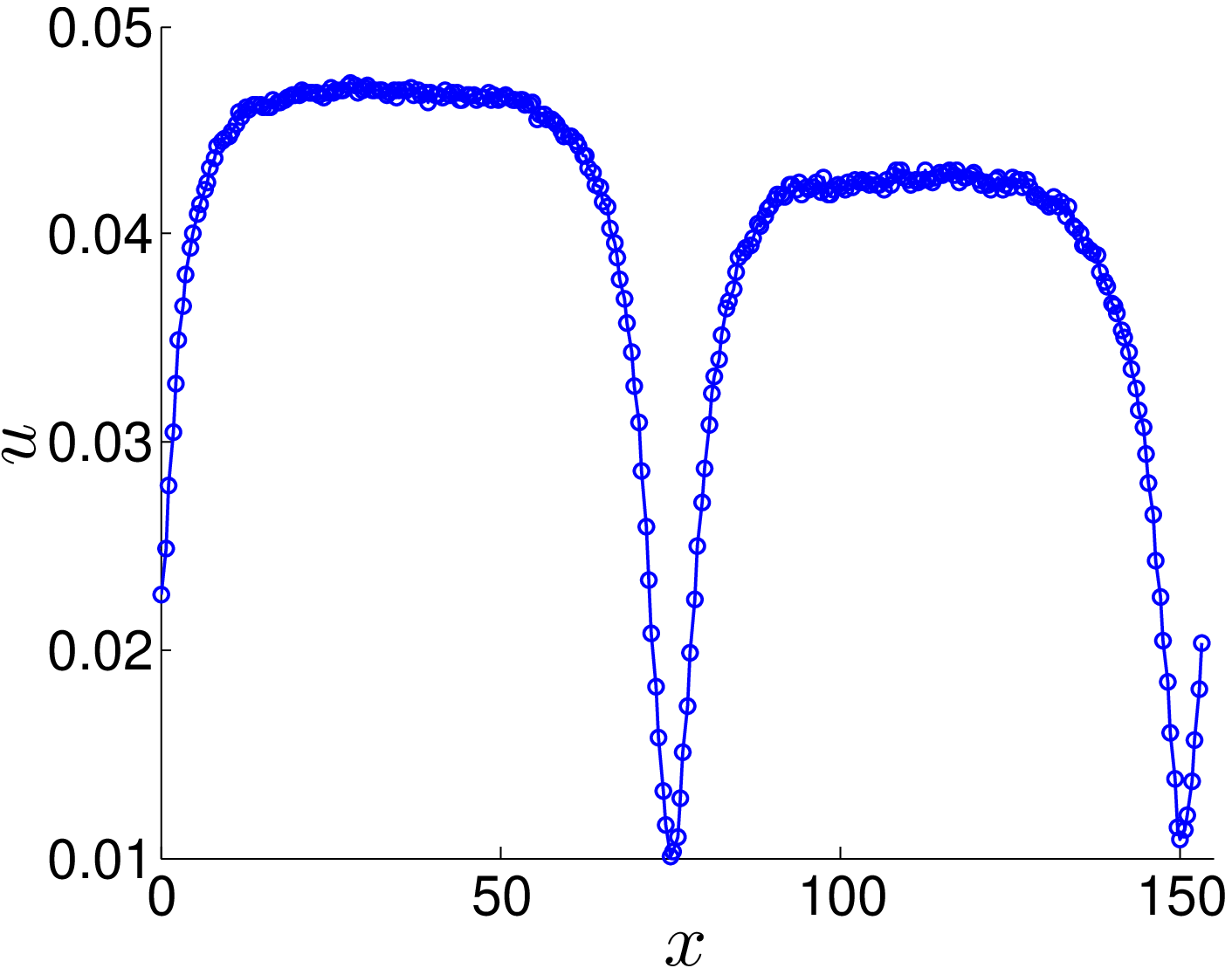}}
  \end{minipage}
 \begin{minipage}{0.49\linewidth}
 \centerline{
  \includegraphics[width=0.85\textwidth]{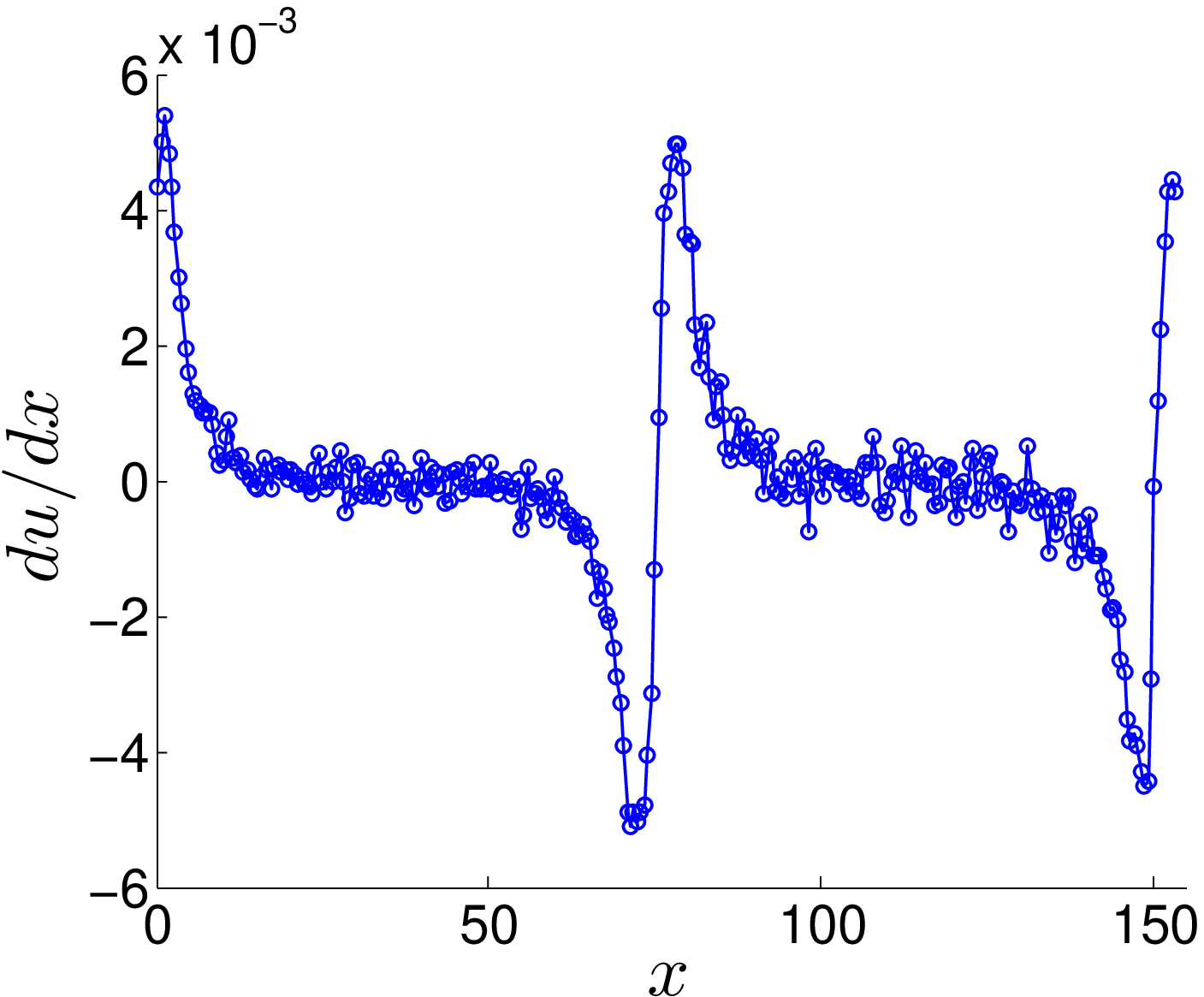}}
 \end{minipage}
   \caption{The distribution of (a) the tangential component of fluid velocity $u$ and (b) the tangential component of linear strain rate ($du/dx$) at the top wall.}
 \label{fig:Vel_StrainRate}
 \end{figure}

\begin{figure}
 \begin{minipage}{0.49\linewidth}\raggedright (a)\end{minipage}
 \begin{minipage}{0.49\linewidth}\raggedright (b)\end{minipage}\\
 \begin{minipage}{0.49\linewidth}
 \centerline{
  \includegraphics[width=0.85\textwidth]{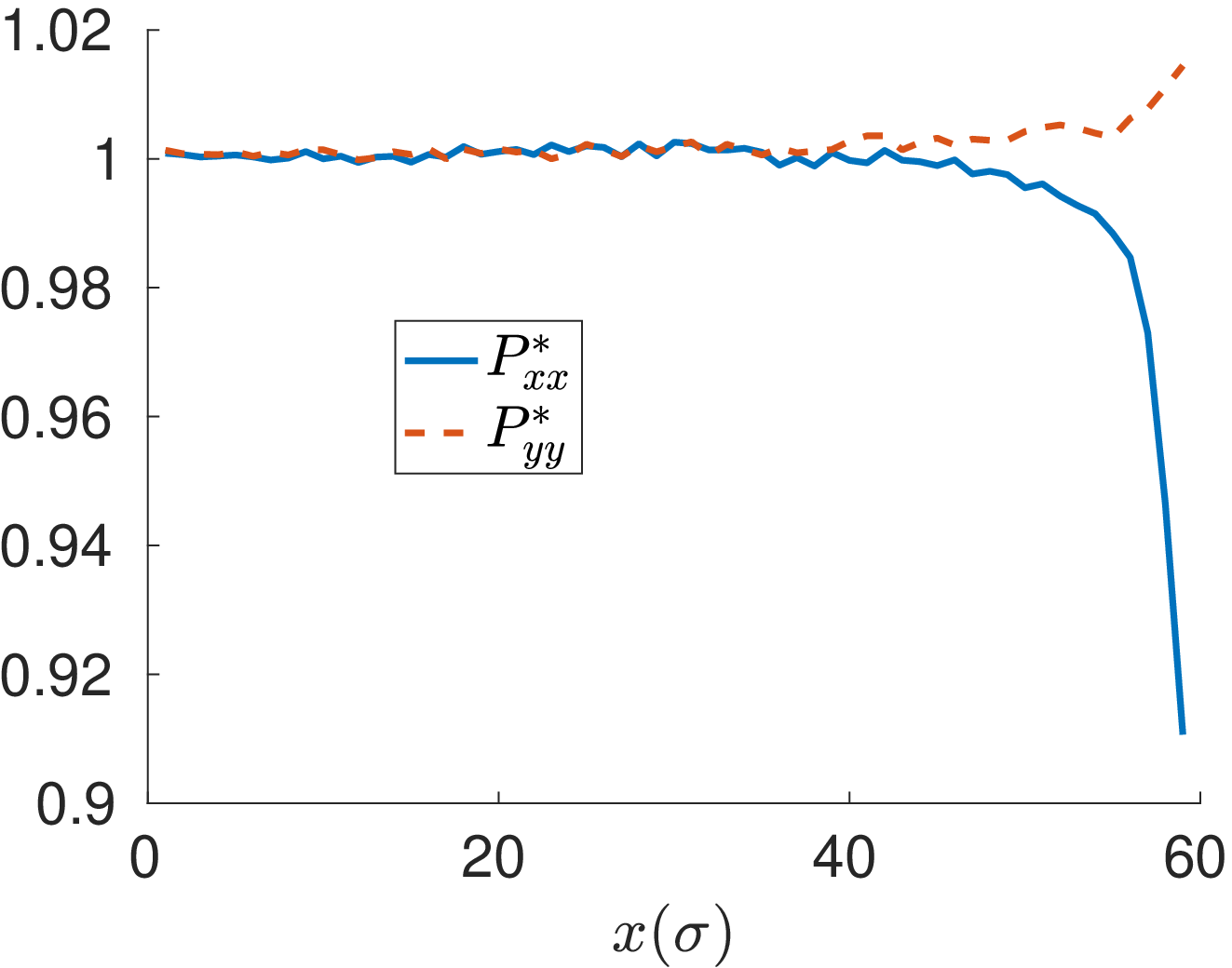}}
  \end{minipage}
 \begin{minipage}{0.49\linewidth}
 \centerline{
  \includegraphics[width=0.85\textwidth]{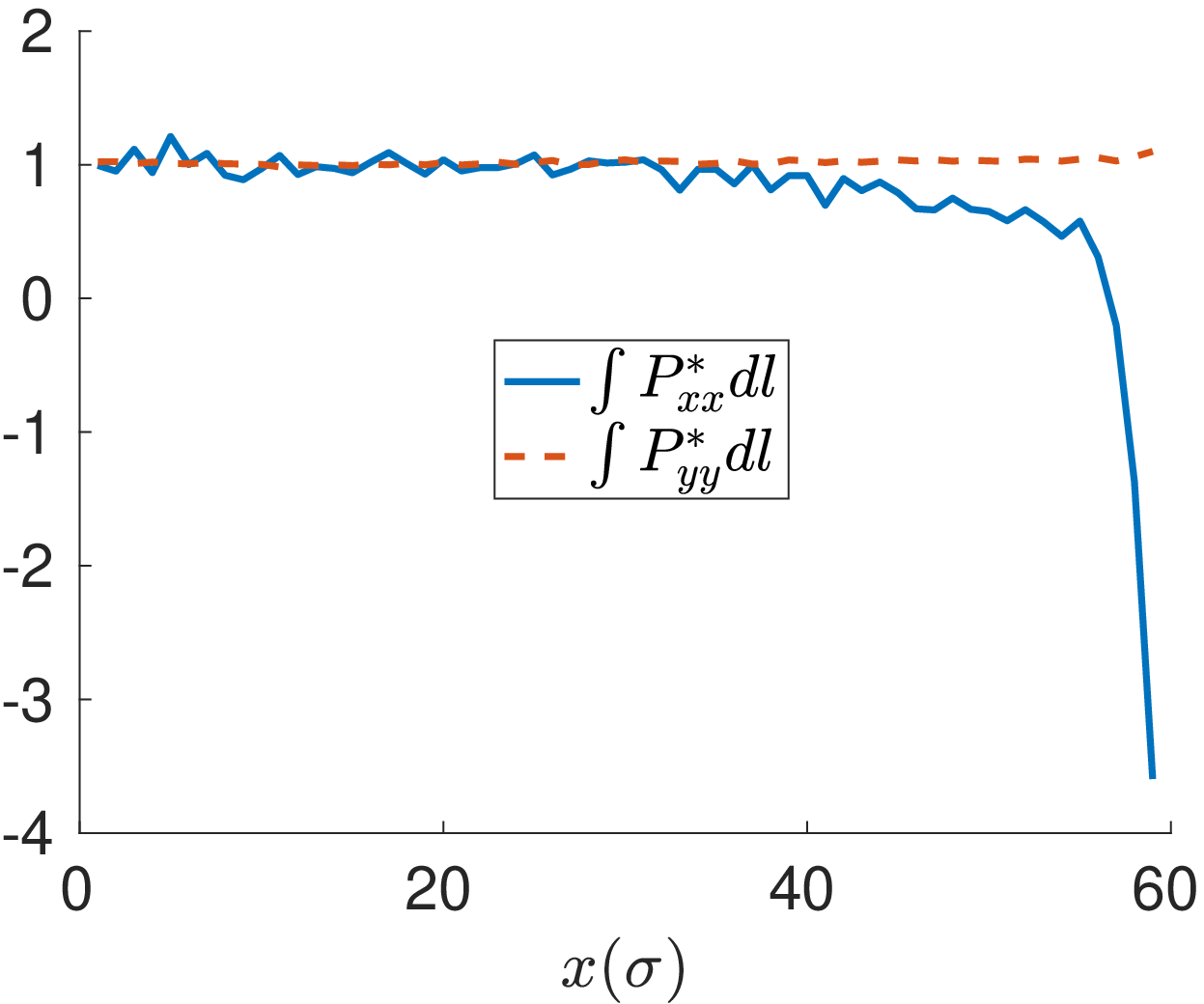}}
 \end{minipage}
   \caption{The distribution of (a) components of linear stress and (b) the cumulative force due to the components of linear stress, defined as $S_{xx}(x) = \int_{H/2}^H P_{xx}(x,y) dy$ and $S_{yy}(x) = \int_{H/2}^H P_{yy}(x,y) dy$ at the top wall.}
 \label{fig:LinearStress}
 \end{figure}

{\it The role and cause of surface tension gradient.} Next, we further investigate the distribution of  surface tension along an interface. Looking at figure \ref{fig:SurfaceTension} it is seen that in the vicinity of the contact line the values for the dynamic cases begin to deviate from the corresponding equilibrium static values. This not only alters the value of surface tension next to the contact point, but also hints that the force contributions due to the surface tension gradient terms in eq. \ref{eq:general_mom_2D} might be relevant. Here, we provide a phenomenological explanation as to the cause of surface tension gradient along the interface, even when there is no temperature or concentration gradient along it. Consider a streamline in the vicinity of the contact point. Assuming the interfaces and the contact point are material boundaries, the streamline on approaching the contact point is forced to turn and hence, decelerate along the direction parallel to the interface. This deceleration results in a linear strain rate tangential to the interface (du/dx, if interface is parallel to $x$ direction). The presence of linear strain rate is further evident on looking at the tangential component of fluid velocity, next to the wall (figure \ref{fig:Vel_StrainRate}(a)). In line with previous numerical \citep{RobbinsMO:01a, Mohseni:16b,QianT:03a} and experimental \citep{BreuerK:03c} findings, a sharp increase in slip is observed in the vicinity of the contact point. As a result of slip, there is an increase in the magnitude of the tangential component of linear strain rate ($du/dx$) near the moving contact line (figure \ref{fig:Vel_StrainRate}(b)). The tangential component of linear strain rate can also be referred to as the component of spatial or convective acceleration along the interface. The occurrence of linear strain rate ($du/dx$) in the vicinity of the contact line, leads to an increase in linear stress ($P_{xx}$) along the interface, see figure \ref{fig:LinearStress}(a). The existence of a gradient in linear strain rate (or convective acceleration) and linear stress, in the vicinity of the contact line, and its importance in accurately defining the slip boundary condition have been discussed by \cite{Mohseni:16b} and \cite{QianT:03a}, respectively. The gradient in linear stress, increases the anisotropy of the stress tensor ($P_{xx}\neq P_{yy}$). Further, we also plot the cumulative force per unit length (per unit length in $z$ direction) along the top wall (figure \ref{fig:LinearStress}(b)), which we define as $S_{xx}(x) = \int_{H/2}^H P_{xx}(x,y) dy$ and $S_{yy}(x) = \int_{H/2}^H P_{yy}(x,y) dy$. Finally, referring back to the mechanical definition of surface tension, $\gamma = \int_{-\infty}^\infty [P_{yy}-P_{xx}]dl$ \citep{Kirkwood:48a}, one can see how the increase in anisotropy caused by the convective acceleration ($du/dx$) results in a gradient in surface tension, in the vicinity of the contact point.
This is further discussed in Appendix B. 

\cite{BlakeTD:95a}, have discussed the existence of surface tension gradients in the vicinity of the contact point with respect to the interface formation theory. They provide two arguments for this occurrence. First, the force balance at the contact point is given by Young's equation, even for the dynamic case. Therefore, since the contact angle no longer corresponds to the static value, then one of the surface tension values must be different. However, the formal derivation of the microscopic contact angle presented here suggests that Young's equation is only valid for the static case and is a special case of the more general contact angle model. Their second argument is based on the theory that a droplet moves by a rolling mechanism. Hence, the particles that once were a part of the liquid-gas interface will eventually become a part of the liquid-solid interface. Therefore, the surface tension of the particle has to gradually change from its value corresponding to the liquid-gas interface to that of the liquid-wall. Our findings suggest that unlike the argument presented by Blake \& Shikhmurzaev, the gradient in surface tension is a result of the convective acceleration caused by the corner formed by the interfaces near the contact line. { It can be said that convective acceleration (or velocity gradient along the interface) is in fact the fundamental source of surface tension gradient. Temperature gradient, concentration gradient and even gradient in charge along the interface, all could result in convective acceleration in the interface, which is in turn responsible for the surface tension gradient. Here, for our case this convective acceleration is a result of varying degree of slip in the vicinity of a moving contact line.}
 
{\it Surface tension gradient's role in determining the leading and trailing contact angles.} The gradient of surface tension, apart from making an important contribution in determining the microscopic dynamic contact angle, also provides an explanation for the difference in contact angle at the leading and trailing edges of a steadily translating droplet. This difference in the contact angles is verified in the results presented in table \ref{tab:contact_angle}, which also include the limiting case of zero wall velocity so that $\theta_1\approx\theta_2$, which is in agreement with Young's equation. This is also consistent with the literature as the wall modeled here is homogeneous and has no macroscopic roughness \citep{McCarthyTJ:06a,McCarthyTJ:10a,LeeTR:13a}. Physically, the difference in contact angles is explained by referring back to figure \ref{fig:SurfaceTension} (a) and table \ref{tab:force}, from which it can be seen that the surface tension along the wall-fluid interface decreases at the trailing edge, while it increases at the leading edge, resulting in a momentum imbalance. Smaller values of the surface tension at the trailing edge indicate an increased local wettability, and hence the fluid-fluid interface tends to form a smaller contact angle. In the same way, at the leading or advancing contact line there is a larger value of the surface tension which results in a larger contact angle. As previously discussed, this is directly attributed to the convective acceleration of the flow along the interface of the wall towards the moving contact line.

 \section{Conclusion}
 In this manuscript following the Gibb's interpretation of an interface, we have systematically derived a model for the microscopic dynamic contact angle. In doing so, we have identified that in addition to surface tension, one of the key parameters is surface tension gradient. This is because, not only does it contribute an additional force but it also is responsible for the local deviation of surface tension value form its static value. Using molecular dynamics simulations the model is validated for different cases of wall-fluid properties and for varying wall velocities. It is shown that in the static limit the model is consistent with Young's equation. In addition, we provide a phenomenological explanation for attributing the surface tension gradient in the vicinity of the contact line, in the absence of temperature or concentration gradients, to convective acceleration. {This is an outcome of varying degree of slip in the vicinity of a moving contact line.} When considering the case of a steadily moving droplet, this surface tension gradient and associated momentum imbalance at the contact line, are found to be the fundamental causes of differing contact angles at the trailing and leading edges of the droplet.
 
Although the force balance was derived with a focus on the contact line problem, the general momentum equation (equation \ref{eq:general_mom}) can be used in various other cases involving geometric or physical irregularities, such as corner flows or vortex sheets at the trailing edge of an airfoil.
 
 \section{Acknowledgment}
 This research was supported by the National Science Foundation and Office of Naval Research. 
 \clearpage
 \newpage
 
 \section*{Appendix A}\label{appA}
The force contributions of respective terms in the contact angle model, \\
$\int_{\Sigma} \left( \rho_a {\bf v^{(\text{a})}\nabla_{(\text{a})}\cdot v^{(\text{a})}} +  \nabla_{(\text{a})}\gamma  +\left[\!\!\left[  {\bf T}\cdot{\bm \xi} \right]\!\!\right]\right) dA
 +\int_{\text{c}}\left[\!\!\left[  \gamma{\bm \nu} \right]\!\!\right]ds
 = 0$, equation \ref{eq:contact_model} are tabulated in Table \ref{tab:force}. By comparing the respective forces, we observe that the force due to surface tension gradient at the wall-fluid interface is comparable to its surface tension values, while that due to inertia and the stress jump across the fluid-fluid interface, is negligible. The force due to inertia is of the {\it O}($10^{-4}$), hence not listed in the table. This  confirms our previous assumption of neglecting the contribution due to inertia and jump in surface stress, taken while simplifying the contact angle model.
 

 \begin{table}
   \begin{center}
 \def~{\hphantom{0}} 
\begin{tabular}{cccccccccc}
 & & &\multicolumn{2}{c}{\underline{Fluid A-Wall}} &\multicolumn{2}{c}{\underline{Fluid B-Wall}} &\multicolumn{3}{c}{\underline{Fluid A-Fluid B}}\\[3pt]
 %
 Case &$U$ & &$F_{\gamma}$ &$F_{\nabla\gamma}$  &$F_{\gamma}$ &$F_{\nabla\gamma}$  &$F_{\gamma}$ &$F_{\nabla\gamma}$ &$F_{\Delta P}$ \\[16pt]
 \multirow{1}{*}{1~} &\multirow{1}{*}{$0.050$~} &~$\theta_1$~ &~$2.80$~ & ~$0.66$~ & ~$4.83$~ & ~$0.83$~ & ~$11.37$~ & ~$0.05$~ & ~$0.05$~ \\
   & &~$\theta_2$~ &~$3.95$~ & ~$0.51$~ & ~$3.37$~ & ~$1.01$~ & ~$11.04$~ & ~$0.00$~ & ~$0.00$~ \\
   
     \multirow{1}{*}{2~} &\multirow{1}{*}{$0.000$~} &~$\theta_1$~ &~$3.32$~ & ~$0.28$~ & ~$4.36$~ & ~$0.10$~  & ~$14.56$~ & ~$0.13$~ & ~$0.03$~ \\ 
   & &~$\theta_2$~ &~$3.15$~ & ~$0.24$~ & ~$4.32$~ & ~$0.11$~ & ~$14.97$~ & ~$0.05$~ & ~$0.05$~ \\   
  
     &\multirow{1}{*}{$0.050$~} &~$\theta_1$~ &~$2.64$~ & ~$0.61$~ & ~$4.91$~ & ~$0.11$~ & ~$10.73$~ & ~$0.04$~ & ~$0.03$~ \\ 
   & &~$\theta_2$~ &~$3.90$~ & ~$0.16$~ & ~$3.77$~ & ~$0.23$~ & ~$10.25$~ & ~$0.03$~ & ~$0.07$~ \\   
  
     \multirow{1}{*}{3~}   &\multirow{1}{*}{$0.000$~} &~$\theta_1$~ &~$2.12$~ & ~$1.85$~ & ~$4.28$~ & ~$0.02$~ & ~$9.55$~ & ~$0.07$~ & ~$0.08$~ \\
   & &~$\theta_2$~  &~$1.94$~ & ~$0.47$~ & ~$4.46$~ & ~$0.57$~ & ~$9.31$~ & ~$0.08$~ & ~$0.11$~ \\
     
     &\multirow{1}{*}{$0.025$~} &~$\theta_1$~ &~$1.40$~ & ~$0.35$~ & ~$4.62$~ & ~$0.20$~ & ~$11.14$~ & ~$0.15$~ & ~$0.07$~ \\
   & &~$\theta_2$~  &~$2.10$~ & ~$0.93$~ & ~$3.99$~ & ~$0.32$~ & ~$10.66$~ & ~$0.07$~ & ~$0.12$~\\
     
     &\multirow{1}{*}{$0.050$~} &~$\theta_1$~ &~$0.93$~ & ~$0.35$~ & ~$4.71$~ & ~$0.29$~ & ~$12.20$~ & ~$0.16$~ & ~$0.08$~ \\
   & &~$\theta_2$~  &~$2.26$~ & ~$0.25$~ & ~$3.70$~ & ~$0.12$~ & ~$9.74$~ & ~$0.16$~ & ~$0.14$~\\
   
   &\multirow{1}{*}{$0.075$~} &~$\theta_1$~ &~$0.57$~ & ~$0.17$~ & ~$4.68$~ & ~$0.15$~ & ~$11.61$~ & ~$0.04$~ & ~$0.09$~ \\
   & &~$\theta_2$~  &~$2.67$~ & ~$0.38$~ & ~$3.51$~ & ~$0.11$~ & ~$8.66$~ & ~$0.02$~ & ~$0.17$~ \\
   \end{tabular}
   \caption{\label{tab:force_contrib} Force contribution due to surface tension, surface tension gradient and pressure jump across respective interfaces are presented. The force contribution due to inertia is of the {\it O}($10^{-4}$) and hence neglected. Here the contact angle is measured with respect to fluid A. As the domain is periodic in $z$ direction we assume the simulation to be two dimensional. Hence, the force presented is in units of force per unit length $(\epsilon\sigma^{-2})$ and the velocity has units of, $\sigma\tau^{-1}$.}
   \label{tab:force}

   \end{center}
 \end{table}

 \section*{Appendix B}\label{appB}
\begin{figure}
   \centerline{\includegraphics[width=0.95\textwidth]{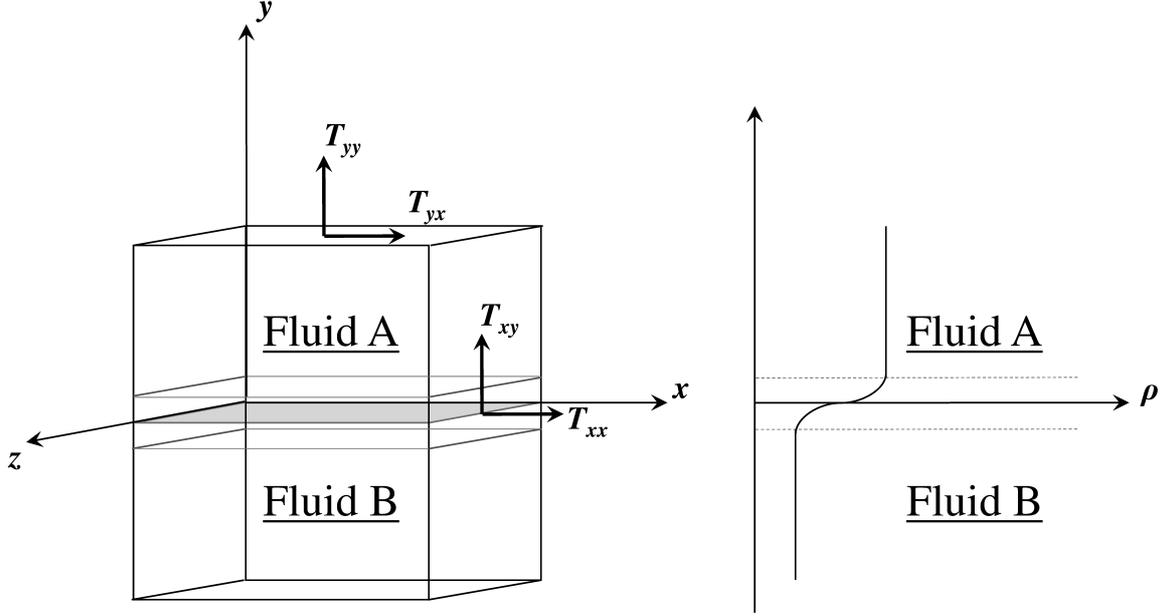}}
   \caption{Schematic of a control volume encompassing a two phase system and a sketch of the corresponding density profile.}
 \label{fig:Schematic_surftension}
 \end{figure}

In order to elaborate on the convective acceleration being the source of surface tension gradient, we start by revisiting the physical meaning and the definition of surface tension. If we consider an infinitesimal element in an isotropic fluid, far away from the interface the components of pressure will be identical, see figure \ref{fig:Schematic_surftension}. If now the element is at the interface, since the forces acting on the element due to fluid A and fluid B are asymmetric, the the components of pressure will no longer be identical. It is this anisotropy in pressure at an interface that constitutes the surface tension. As per the mechanical definition, \cite{Kirkwood:48a} described surface tension as the force in excess of the contribution of uniform normal pressure. Next, we show that if the net force on a control surface due to stresses is systematically computed, such that the control volume encompasses the interface, one naturally obtains the surface tension force. Now, consider a control volume extending from the middle of fluid A to that of fluid B, such that it encompasses the interface. The total stress is denoted by $\boldsymbol{T}$. The net force due to surface stresses can be written as, 
\begin{equation}
 \boldsymbol{F}_{\text{surf}} = \int_{\Sigma} \boldsymbol{T \cdot dA}.
\end{equation}
It must be noted that typically $\boldsymbol{T}$ is considered to be continuous, but here it is a discontinuous function as it cuts across the interface. Hence, when we convert the area integral to a volume integral using the surface divergence theorem we obtain
\begin{equation}
 \boldsymbol{F}_{\text{surf}} = \int_{R} \boldsymbol{\nabla T \cdot dV} + \int_{\Sigma} \left[\!\!\left[\boldsymbol{T \cdot \xi} \right]\!\!\right] dA.
\end{equation}

This jump in stress across the interface is equal to the surface stress. Therefore, if we assume that dynamical properties of the interface are negligible, then the surface stress in turn is equivalent to the surface tension \citep{LealLJ:07a}, $ \int_{\Sigma} \left[\!\!\left[\boldsymbol{T \cdot \xi} \right]\!\!\right] dA = \int_{c} \gamma \nu dl$, and hence 
\begin{equation}
 \boldsymbol{F}_{\text{surf}} = \int_{R} \boldsymbol{\nabla T \cdot dV} + \int_{c} \gamma \nu dl.
\end{equation}
Hence, by allowing the fluid stress to include a discontinuity, we have shown that the matching condition for surface tension is naturally obtained.
 
%
%
%
%
%

\bibliographystyle{jfm}

\bibliography{RefA2}

\end{document}